\documentclass[12pt]{article}
\usepackage{graphicx}
\usepackage{appendix}

\begin{document}

\title{Influence of surface diffusion on catalytic reactivity of spatially
inhomogeneous surfaces -- mean field modeling}

\maketitle

\noindent
\textbf{Lukasz Cwiklik} \footnote{Presently at Fritz Haber Institute
for Molecular Dynamics, Hebrew University, Jerusalem 91904, Israel;
email: cwiklik@gmail.com}\\ Institute of Organic Chemistry and
Biochemistry, Academy of Sciences of the Czech Republic, and Center
for Biomolecules and Complex Molecular Systems, Flemingovo nam. 2,
16610 Prague 6, Czech Republic\\

\begin{abstract}
Kinetics of model catalytic processes proceeding on inhomogeneous
surfaces is studied. We employ an extended mean-field model that takes
into account surface inhomogeneities. The influence of surface
diffusion of adsorbent on the kinetics of the catalytic process is
investigated. It is shown that diffusion is responsible for
differences in the reaction rate of systems with different
arrangements of active sites. The presence of cooperative effects
between inactive and active sites is demonstrated and the conditions
when these effects are important are discussed. We show that basic
catalytic phenomena on nonuniform surfaces can be studied with
mean-field modeling methods.
\end{abstract}

\section{Introduction} \label{sec:intro}

Catalytic processes on spatially inhomogeneous surfaces are examples
of heterogeneous reacting systems where the heterogeneity means not
only that a reaction takes place on the interface but also that the
catalyst surface is heterogeneous. Current advances in nanotechnology
made it possible to produce nanopatterned surfaces with assumed
geometric properties and to study their reactivity. In the case of
these surfaces the heterogeneity occurs in the nanometer scale and in
the presence of finite surface diffusion may strongly influence the
kinetics of surface processes and hence play a key role in the surface
reactivity \cite{sam2006, zhdanov2002}. A proper description of the
kinetics on nanostructured surfaces is, in most cases, impossible
using only the standard phenomenological chemical rate equations
\cite{temel2007, chonkbook}. The classical description assumes an
ideal mixing of adsorbate particles and excludes the presence of both
the concentration gradients and the spatial nonuniformities of the
surface. To overcome these limitations two groups of methods are
usually employed in theoretical studies: kinetic Monte Carlo
simulations (KMC) where adsorbate particles can directly undergo all
studied surface processes and extended mean-field modeling (MF) where
spatial inhomogeneities are taken into account indirectly. In most of
the former studies surface diffusion of adsorbate was either neglected
or considered only as a rapid process. Even for uniform catalysts
surfaces diffusion was neglected in early works, with the well-known
Ziff-Gulari-Barshad model as an example \cite{ziff1986}. In the case
on nonuniform surfaces in early works, representing both MC and MF
treatment, surface diffusion was not included explicitly
\cite{jansen1999} or even completely neglected \cite{hermse2000,
mcleod2000, barnes2001}. The kinetics was also studied in the limit of
rapid diffusion where local nonuniformities were not included
\cite{tammaro1995, james1999, evans2002, liu2002, mastny2006}. In
recent works concerning nonuniform surfaces it is usual to include
surface diffusion in MC treatment \cite{zhdanov2002, caballero2005,
cwiklik_spill, hu2005} whereas in MF models only approximate
descriptions were employed (diffusion between two regions -- see
\cite{zvejnieks2002}) or a process with coverage dependent diffusion
coefficient). However, these MF models were dealing only with uniform
surfaces \cite{caballero2005}. Evans and co-workers investigated the
role of finite surface diffusion on the poisoning of catalyst surfaces
using MF methods but also for uniform surfaces \cite{tammaro1998,
liu2006}. In this communication an extended mean-field description
based on the Master equation formalism and describing the kinetics of
catalytic processes (including finite surface diffusion) on
heterogeneous surface is presented. This formalism takes into account
both a local arrangement of reactive and non-reactive surface sites
and local arrangement of adsorbate. Therefore, it lets to include and
investigate an influence of geometric surface heterogeneities on the
kinetics of considered catalytic processes in the regime of finite
surface diffusion.

\section{Mean-field model}
\label{sec:model}

\begin{figure} [!hbp]
 \centering \includegraphics[width=0.32\textwidth]{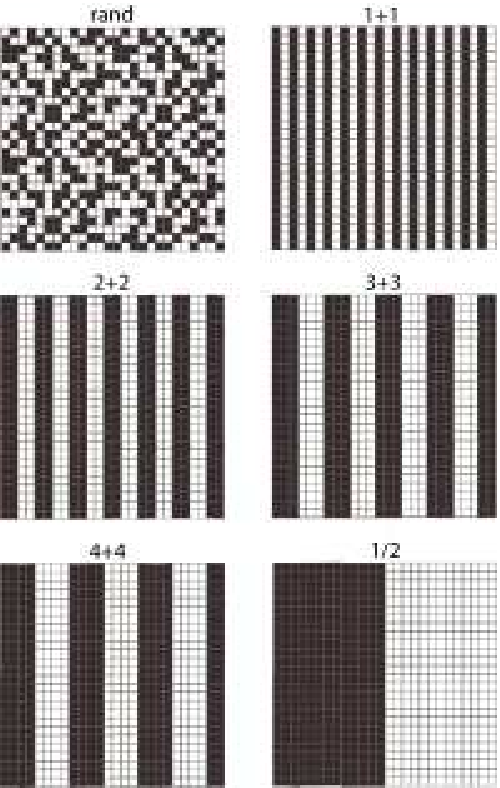}
 \caption{Schematic representations of considered surfaces, white and
  gray lattice sites represent, respectively, inactive and active
  surface centers.}
 \label{fig:schem_latt}
\end{figure}

The mean-field model introduced here was derived by averaging Master
equations describing a time evolution of the density of the
probability of occurring different local adsorbate/surface
configurations on the catalyst surface. A similar method was proposed
previously by other authors, however, they did not study in detail the
role of finite surface diffusion for non-uniform surfaces
\cite{jansen1999, hermse2000, dickman1986}. We employed a coarse grain
model of a surface with a surface represented as a two dimensional
lattice. In order to study non-uniform surfaces we considered 2 types
of lattice sites: active and inactive in the surface
reaction. Adsorption, desorption and diffusion were allowed to proceed
at each site, and reaction only on the reactive ones. We were studying
surfaces with the increasing ordering in the active sites
distribution, starting from the random distribution, then active sites
were arranged in the form of stripes of increasing width (see
Fig. \ref{fig:schem_latt}). Two model catalytic reactions were
considered: unimolecular and bimolecular (the latter according to
Langmuir-Hinshelwood mechanism). First we investigated the catalytic
process with an unimolecular reaction of the type $A \rightarrow
P$. At a surface with two types of centers the process can be
presented as:
\begin{eqnarray} 
A_{\left(g\right)} + o & \rightleftharpoons & A_{\left(ads\right)} \label{eq:ads_on_inact}
\\
A_{\left(g\right)} + * & \rightleftharpoons & A^{*} \label{eq:ads_on_act}
\\
A_{\left(ads\right)} + * & \rightleftharpoons & A^{*} + o \label{eq:surf_diff}
\\
A^{*} & \rightarrow & P_{\left(g\right)} + * \label{eq:surf_react}
\end{eqnarray}
where: $A_{\left(g\right)}$ --~reactant particle in the gas phase, $o$
-- surface center inactive in the surface reaction, $A_{(ads)}$ --
reactant particle adsorbed on an inactive surface center, $*$ --
active surface center, $A^*$ --~reactant particle adsorbed on an
active center, $P_{\left(g\right)}$ --~product particle desorbed to
the gas phase.  Eqs. (\ref{eq:ads_on_inact}) and (\ref{eq:ads_on_act})
describe reversible adsorption of reactant on, correspondingly,
inactive and active surface centers. Eq. (\ref{eq:surf_diff})
corresponds to surface diffusion of reactant particles between active
and inactive surface centers. Eq. (\ref{eq:surf_react}) depicts the
unimolecular reaction proceeding on active centers and the immediate
desorption of product to the gas phase (the assumption of immediate
desorption of product particles is widely used in modeling of surface
reactions, for example, in Ziff-Gulari-Barshad model; generally,
product particles may stay or diffuse on the surface and poison the
catalyst, however, for many practically important catalysts, product
desorption is much more efficient than other surface processes and can
be considered as a rapid phenomenon). This model includes information
about the surface heterogeneity but only in the form of spatially
averaged concentrations of reactant on both active and inactive
centers.

In order to describe the local structure of the surface we considered
probabilities of occurring different one-site configurations:
$\left[\circ\right]$ -- probability of occuring an inactive unoccupied
site, $\left[\ast\right]$ -- active unoccupied, $\left[A\circ\right]$
-- inactive occupied and $\left[A\ast\right]$ -- active occupied. We
took into account all processes of creation and annihilation of these
configuration including surface diffusion (for instance, an inactive
unoccupied site can be generated either by desorption of reactant
particle from an occupied inactive site or by diffusion of a particle
from an occupied inactive site onto an unoccupied site neighboring to
the considered one). The following set of equations describes a time
evolution of one-site probabilities:
\begin{eqnarray} 
  {d[\circ] \over dt} & = & P_{des,inact} [A\circ] +
  P_{diff,IA} [A\circ|*] -\nonumber
  \\
  & & - P_{ads,inact} [\circ] - P_{diff,AI} [A*|\circ]
  \label{eq:master_o}
  \\
  {d[*] \over dt} & = & P_{des,act} [A*] + P_{diff,AI} [A*|\circ] +
  P_r [A*] -\nonumber
  \\
  & & - P_{ads,act} [*] - P_{diff,IA} [A\circ|*]
  \label{eq:master_star}
  \\
  {d[A*] \over dt} & = & P_{ads,act} [*] + P_{diff,IA} [A\circ|*]
  -\nonumber
  \\
  & & - P_{des,act} [A*] - P_{diff,AI} [A*|\circ] - P_r [A*]
  \label{eq:master_a}
  \\
  {d[A\circ] \over dt} & = & P_{ads,inact} [\circ] + P_{diff,AI}
  [A*|\circ] -\nonumber
  \\
  & & - P_{des,inact} [A\circ] - P_{diff,IA} [A\circ|*]
  \label{eq:master_ao}
\end{eqnarray}
where: $P_{i,j}$ -- probabilities (per time unit) of
desorption/adsorption of reactant particle on the inactive/active
sites, $P_{diff,AI}$ ($P_{diff,IA}$) -- probability of diffusion of A
particle from an active (inactive) onto an inactive (active) site,
$P_{r}$ -- probability of surface reaction. On right-hand sides of
above equations there are probabilities of occurring of two-center
lattice configurations which are equal to the probabilities of finding
in the system the pairs of neighboring sites in the given states. For
example, the term $[A\circ|*]$ depicts the probability of occurring
the local configuration where an inactive occupied lattice site has an
active unoccupied site in the nearest neighborhood. The changes of
two-center terms which are present on the right-hand side of these
equations can be written as a function of three-center probabilities:
\begin{equation} 
  {d[i|j] \over dt} = f\left([k|l], [m|n|o]\right)
\end{equation}
The simplest way to solve such a hierarchy of equations is to use
one-center approximation, where $[i|j] \approx [i][j]$. However, this
one-center description omits the presence of spatial correlations. It
can be noticed that two-center configurations characterized by
probabilities $\left[A\ast|\circ\right]$ and
$\left[A\circ|\ast\right]$ may be found on the surface only if active
and inactive sites neighbor each other and the presence of such
configurations depends on the spatial arrangement of active
centers. For example, in the case of randomly placed active centers
with a low value of $\theta_{act}$ (ratio of active to inactive sites)
almost each active site on the lattice has inactive neighbors. On the
other hand, if spatial correlations in the sites arrangement are
present, a significant fraction of active sites may have no inactive
neighbors. Therefore, spatial correlations in the arrangement of active
centers can be included by introduction of an 'inhomogeneity
parameter' to approximate two-center probabilities:
\begin{eqnarray}
  \left[A\ast|\circ\right] & \approx & w [A\ast][\circ] \label{eq:factorization_uni_1}
  \\
  \left[A\circ|\ast\right] & \approx & w [A\circ][\ast] \label{eq:factorization_uni_2}
\end{eqnarray}
where the inhomogeneity parameter $w$ is an average probability that
an active center neighbors an inactive one in the considered system
(it depends on the number of active-inactive neighbors in the system)
and $w$ can be calculated for each arrangement of active centers. In
this approximation only one-center concentration terms are explicitly
present, however, information about two-center correlations in the
arrangement of surface sites is also taken into account with the
inhomogeneity parameter. Similar approximation was proposed by Jansen
and Hermse \cite{jansen1999}. We obtained an analytical solution for
stationary states of eqs. (\ref{eq:master_o}) -- (\ref{eq:master_ao})
with the approximation given by eqs. (\ref{eq:factorization_uni_1})
and (\ref{eq:factorization_uni_2}) taking an additional assumption
that the probability of diffusion between active and inactive regions
is equal, i.e., $P_{diff,AI} = P_{diff,IA}$ (for analytical solutions
see Appendices).

We also studied the kinetics of bimolecular surface reaction of the
type: $2A \rightarrow P$ proceeding according to Langmuir-Hinshelwood
scheme. In this case eq. (\ref{eq:surf_react}) has the form: $2A^{*}
\rightarrow P_{\left(g\right)} + 2 *$ and two-center terms
$\left[A\ast|A\ast\right]$ appear. We approximated them introducing an
inhomogeneity parameter $w_r$ which corresponds to the probability of
occurring at least one active center in the nearest vicinity of
considered active site. Assuming, as previously, $P_{diff,AI} =
P_{diff,IA}$ we obtained analytical solution for stationary states
(see Appendices).

\section{Results}
\label{sec:results}

During this study we focused on the influence of both the arrangement
of surface sites and surface diffusion on the kinetics of catalytic
process. Therefore, we assumed $P_{ads,act} = P_{ads,inact} = p_{ads}$
and $P_{des,act} = P_{des,inact} = p_{des}$ (i.e., all surface sites
had the same adsorption/desorption properties) and $P_{diff,AI} =
P_{diff,IA} = p_{diff}$ (due to the conditions for our analytical
solutions). In order to test the mean-field approach presented above
we compared mean-field results with the results obtained in our
earlier work for similar system where Monte Carlo simulations were
used \cite{cwiklik_ssci}. Both surface coverages and turn-over numbers
(results not presented here) were in a very good quantitative
agreement with MC results in the case of unimolecular reaction and in
a good qualitative agreement for bimolecular mechanism (a comparison
of these results is provided in Appendices). Particularly, the MF
model was able to reproduce MC results distinguishing between surfaces
with different arrangements of active sites. Therefore, we employed
the mean-field approximation with inhomogeneity parameters in the
further calculations.

\begin{figure} [!hbp]
 \centering
 \includegraphics[width=0.4\textwidth]{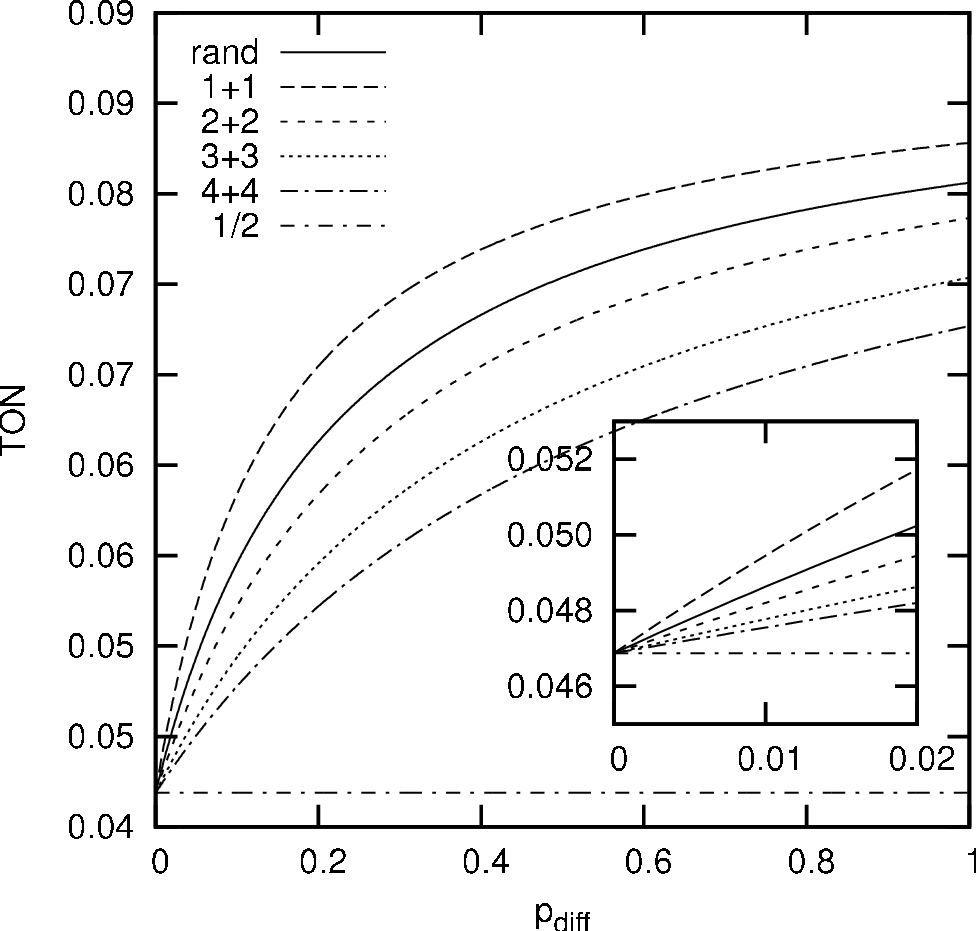}
 \caption{Turn-over number vs. probability of surface diffusion for
 unimolecular reaction ($\theta_{act} = 0.5$, $x_{gas} = 0.05$,
 $p_{des} = 0.01$, $p_r = 0.9$).}
 \label{fig:uni_ton_pdiff}
\end{figure}

Fig. \ref{fig:uni_ton_pdiff} presents turn-over number as a function
of surface diffusion probability for surfaces with different
arrangements of active sites in the case of unimolecular reaction for
half of centers being active. The arrangement of active sites strongly
influences catalytic activity of the system resulting in different TON
values for different types of the arrangement. As the ordering of
active sites increases (in a series: '1+1', '2+2', '3+3', '4+4',
'1/2') the values of TON decrease. The random arrangement leads to TON
values located between these for '1+1' and '2+2'. For low values of
$p_{diff}$ the differences in TON are relatively small because surface
diffusion is slow and the adsorption/desorption balance is the main
factor influencing the rate of the overall process. As $p_{diff}$
increases the differences become more pronounced and '1+1' and 'rand'
systems are clearly the most efficient ones. Particularly important is
the case of '1/2' arrangement where TON is constant over the whole
$p_{diff}$ range and TON values are the lowest. In each case half of
surface centers are inactive in the reaction, however, they are active
in the adsorption, desorption and diffusion processes. Inactive sites
in the presence of surface diffusion constitute an additional channel
transporting reagent molecules from the gas phase onto active sites
and thus cooperative effects are observed. Similar cooperative effects
between active and inactive surface centers are observed, for example,
for supported metal catalysts \cite{zhdanov2002, cwiklik_spill}. As
the ordering of active sites arrangement increases, the amount of
active sites neighboring inactive ones decreases and this additional
diffusive transport becomes less efficient. For '1/2' system this
transport is negligible, since the length of the border between
active/inactive asymptotically goes to zero.

\begin{figure} [!hbp]
 \centering
 \includegraphics[width=0.4\textwidth]{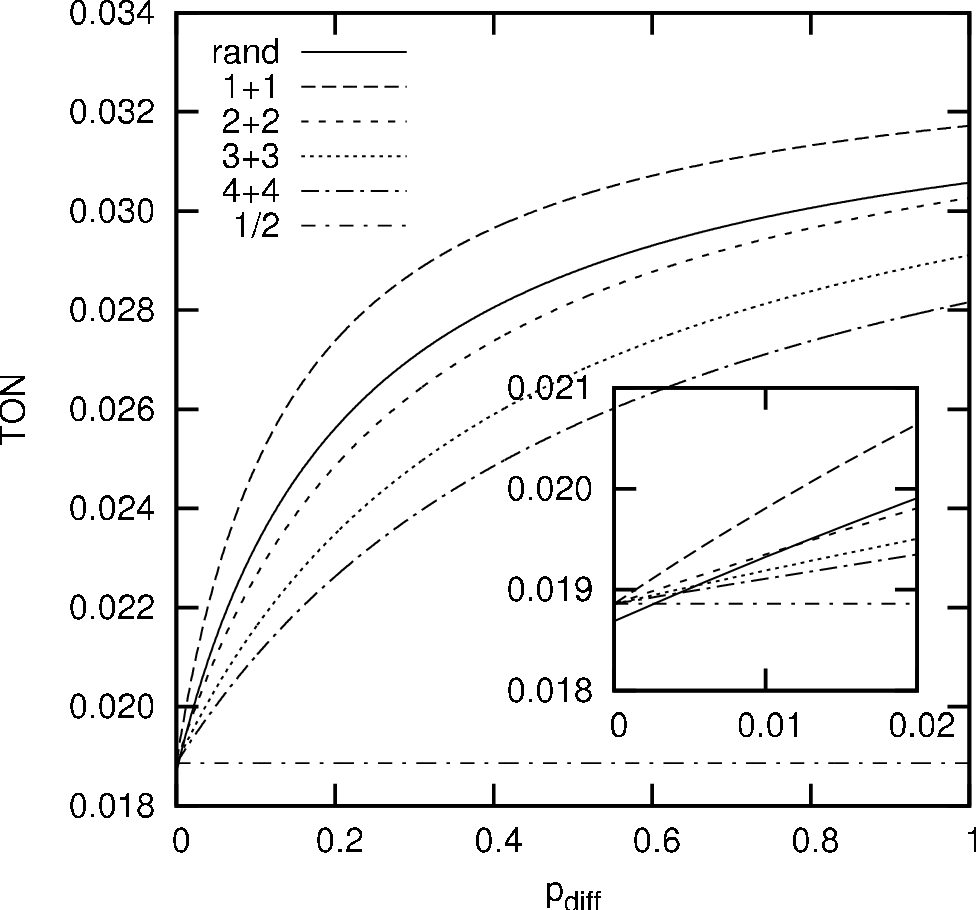}
 \caption{Turn-over number vs. probability of surface diffusion for
 bimolecular reaction ($\theta_{act} = 0.5$, $x_{gas} = 0.05$,
 $p_{des} = 0.01$, $p_r = 0.9$).}
 \label{fig:bi_ton_pdiff}
\end{figure}

In Fig. \ref{fig:bi_ton_pdiff} the values of turn-over number
vs. diffusion probability for bimolecular mechanism are
presented. Here TON is lower than for the unimolecular case, however,
qualitatively this plot resembles unimolecular case with the exception
of both very low $p_{diff}$ values. For bimolecular reaction, the
presence of pairs (at least) of neighboring active sites is a
necessary condition for surface reactivity. For $p_{diff} \rightarrow
0$ (see inset) TON values for all systems, except 'rand', are located
in the same point. For all these systems there exist a relatively high
number of active-active neighbors whereas in the 'rand' case this
number is lower and hence the latter surface is less reactive. As
surface diffusion becomes more significant 'rand' system is more
reactive and above $p_{diff} = 0.012$ it is the second reactive
one. For $p_{diff} < 0.007$ ('rand' more efficient than '1/2') TON
values are determined mainly by the number of active-active neighbors
whereas above this threshold the cooperative effects between active
and inactive sites become more important and under these conditions
active-inactive neighborhoods are also promoting the reaction. This
cooperative effects are observed for all systems except, like in the
unimolecular case, '1/2' one where the influence of surface diffusion
is neglegible.

\begin{figure} [!hbp]
 \centering
 \includegraphics[width=0.6\textwidth]{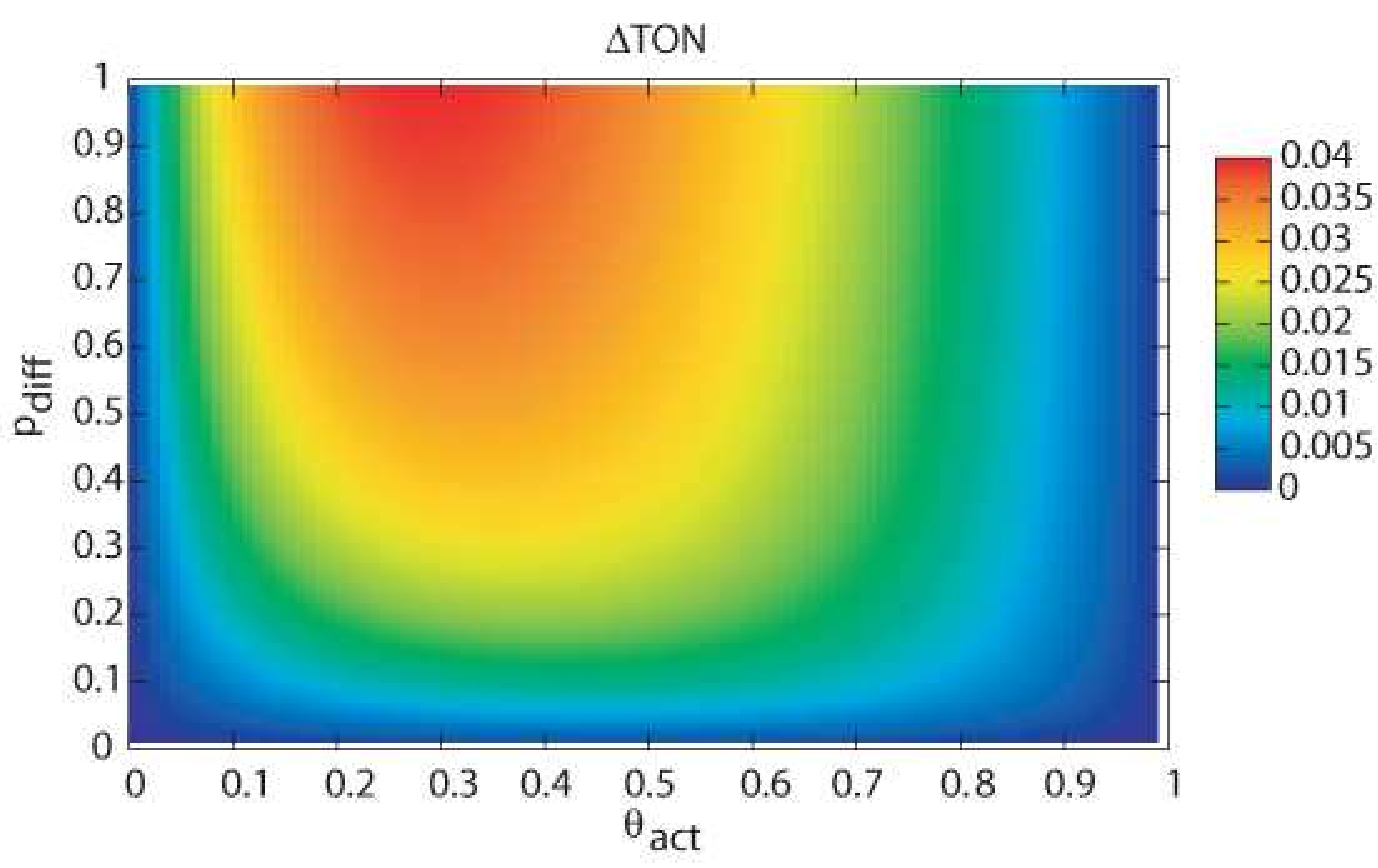}
 \caption{$\Delta$TON vs. fraction of active sites and surface
  diffusion rate for the unimolecuar surface reaction mechanism
  ($x_{gas} = 0.05$, $p_{des} = 0.01$, $p_r = 0.9$).}
 \label{fig:theta_act_p_diff}
\end{figure}

In Fig. \ref{fig:theta_act_p_diff} a difference between TON of 'rand'
and '1/2' system ($\Delta$TON) vs. both $\theta_{act}$ and $p_{diff}$
is presented. This quantity depicts the difference in catalytic
activity of the system with active sites spreaded randomly among
inactive ones and the surface with active sites accumulated in one
surface region. For low $p_{diff}$ there is no influence of the active
sites arrangement on the rate of catalytic process ($\Delta$TON
$\approx 0$). As $p_{diff}$ increases, the difference between 'rand'
and '1/2' system becomes more pronounced and 'rand' system (the one
with the possibility of cooperation between inactive and active sites)
becomes more effective. There is an optimal value $\theta_{act}
\approx 0.32$ for which $\Delta$TON achieves a maximum. At low
$\theta_{act}$, in both 'rand' and '1/2' system, TON is near zero,
therefore, the values of $\Delta$TON are also low. While $\theta_{act}$
becomes higher, the cooperative effects in 'rand' system no longer
influence the catalytic process (the number of inactive sites is very
low) and hence there is no distinction between both systems. We
observed similar behavior in the case of bimolecular reaction
mechanism (results not presented here), in that case the maximum of
$\Delta$TON was located at $\theta_{act} \approx 0.45$. These results
support the above conclusions about the role of surface diffusion in
cooperative effects between active and inactive sites.

\section{Conclusions}
\label{sec:conclusions}

In this work we investigated theoretically the role of surface
diffusion on the chemical kinetics for processes catalyzed by
inhomogeneous surfaces. We employed extended mean-field model which
takes into account the arrangement of active sites on the surface and
we found analytical solutions for two model processes on different
surfaces. We demonstrated that surface diffusion strongly influences
the kinetics and that diffusion is responsible for the differences in
the reactivity of systems with different arrangements of active
sites. The presence of cooperative effects between inactive and active
centers was demonstrated. Inactive sites, due to the possibility of
adsorption and surface diffusion of particles on them, constitute an
additional channel transporting reagent particles onto active
sites. The important message of this study is that these phenomena can
be described with mean-field modeling. It should be stressed that the
results presented here should be treated qualitatively. This study
shows the general features of catalytic systems with inhomogeneous
surfaces and the possible phenomena which may be observed and we
focused solely on the role of the arrangement of surface sites in the
presence of finite surface diffusion. It should be noted that the mean
field description presented here is able to reproduce qualitatively,
for the considered uni- and bimolecular processes, the whole variety
of phenomena (such as spillover/reverse spillover, capture zone
effects, support interactions and communication effects) that are
observed experimentally for more complex systems (the reader can find
a good summary of such experimental results in the review of Libuda
and Freund \cite{libuda2005}).

\appendix
\appendixpage

\section{Solutions of probabilities evolution equations}

Introducing: $[A\ast] = x$, $[A\circ] = y$, $P_{ads,act} = k_1$,
$P_{des,act} = k_2$, $P_{ads,inact} = k_3$, $P_{des,inact} = k_4$,
$P_{diff,IA} = k_5$, $P_{diff,AI} = k_6$, $P_{r} = k_7$, $x_{act} =
a$, $x_{inact} = b$\\ the following set of nonlinear equations can be
written for unimolecular mechanism in the place of
Eqs. (\ref{eq:master_a}) i~(\ref{eq:master_ao}):
\begin{eqnarray} {dx \over dt} & = & k_1 \left(a - x\right) + k_5 w y
  \left(a - x\right) -\nonumber
  \\
  & & - k_2 x - k_6 w x \left(b - y\right) - k_7 x  \label{eq:dyn_1}
  \\
  {dy \over dt} & = & k_3 \left(b - y\right) + k_6 w x \left(b -
    y\right) -\nonumber
  \\
  & & - k_4 y - k_5 w y \left(a - x\right)  \label{eq:dyn_2}
\end{eqnarray}

For bimolecular mechanism we get:

\begin{eqnarray} {dx \over dt} & = & k_1 \left(a - x\right) + k_5 w y
  \left(a - x\right) -\nonumber
  \\
  & & - k_2 x - k_6 w x \left(b - y\right) - 2 k_7 w_r x^2  \label{eq:bi_dyn_1}
  \\
  {dy \over dt} & = & k_3 \left(b - y\right) + k_6 w x \left(b -
    y\right) -\nonumber
  \\
  & & - k_4 y - k_5 w y \left(a - x\right)  \label{eq:bi_dyn_2}
\end{eqnarray}

Here we are interested in the stationary solutions of one-site
probabilities evolution equations (we are going to study steady-state
kinetics) hence we assume:
\begin{equation}
  {dx \over dt} = 0 \mbox{ and } {dy \over dt} = 0
\end{equation}
An analytical solutions for different values of parameter $w$ can be
found assuming that the probability of diffusion between active and
inactive region is equal, i.e., $k_5 = k_6$. Below are the solutions
for uni- and bimolecular mechanism.

\subsection{Unimolecular mechanism}

\begin{eqnarray}
x & = & Z^{-1} \left(\left(a\,b\,k_3+a^2\,k_1\right)\,k_{5}\,w+a\,k_1
    \,\left(k_4+k_3\right)\right) \label{eq:uni_solution_x}
\\
y & = & Z^{-1} \left(\left(b^2\,k_3+a\,b\,k_1\right)\,k_{5}\,w+b\,k_3\,k_7+b\,k_2\, k_3+b\,k_1\,k_3\right) \label{eq:uni_solution_y}
\end{eqnarray}
where:
\begin{eqnarray}
Z & = & \left(a\,k_{5}\,k_7+\left(b\,\left(k_4
 +k_3\right)+a\,k_2+a\,k_1\right)\,k_{5}\right)\,w+ \nonumber
\\
& & \left(k_4+k_3\right)\,k_7+k_2\,\left(k_4+k_3\right)+k_1\,\left(k_4+k_3\right) \label{eq:uni_solution_z}
\end{eqnarray}

\subsection{Bimolecular mechanism}

  In the case of bimolecular reaction mechanism two solutions of the
  set of Eqs. (\ref{eq:bi_dyn_1}) and (\ref{eq:bi_dyn_2}) were
  obtained with opposite sign of $x$. The one with $x>0$ was taken
  into account (we do not present the solution for $y$ here since it
  was not used during TON calculations):
\begin{eqnarray*}
x =&& (((8*a^2*b*k_3+8*a^3*k_1)*k_5^2*k_7*w^2+ \nonumber \\
   && ((8*a*b*k_3+16*a^2*k_1)*k_4+8*a*b*k_3^2+16*a^2*k_1*k_3)*k_5*k_7*w+ \nonumber \\
   && (8*a*k_1*k_4^2+16*a*k_1*k_3*k_4+8*a*k_1*k_3^2)*k_7)*w_r+ \nonumber \\
   && (b^2*k4^2+(2*b^2*k_3+2*a*b*k_2+2*a*b*k_1)*k_4+b^2*k3^2+ \nonumber \\
   && (2*a*b*k_2+2*a*b*k_1)*k_3+a^2*k_2^2+2*a^2*k_1*k_2+a^2*k_1^2)*k_5^2*w^2+ \nonumber \\
   && ((2*b*k_2+2*b*k_1)*k_4^2+((4*b*k_2+4*b*k_1)*k_3+2*a*k_2^2+4*a*k_1*k_2+ \nonumber \\
   && 2*a*k_1^2)*k_4+(2*b*k_2+2*b*k_1)*k_3^2+ \nonumber \\
   && (2*a*k_2^2+4*a*k_1*k_2+2*a*k_1^2)*k_3)*k_5*w+ \nonumber \\
   && (k_2^2+2*k_1*k_2+k_1^2)*k_4^2+(2*k_2^2+4*k_1*k_2+2*k_1^2)*k_3*k_4+ \nonumber \\
   && (k_2^2+2*k_1*k_2+k_1^2)*k_3^2)^{1/2}+(-b*k_4-b*k_3-a*k_2-a*k_1)*k_5*w+ \nonumber \\
   && (-k_2-k_1)*k_4+(-k_2-k_1)*k_3)/ \nonumber \\
   && ((4*a*k_5*k_7*w+(4*k_4+4*k_3)*k_7)*w_r)
\end{eqnarray*}

\section{Comparison with Kinetic Monte Carlo results}

In order to test the usefulness of the presented mean field model it
was employed to describe the kinetics of both a uni- and bimolecular
catalytic process on the model surfaces with various arrangements of
active centers. We started with the lattice models of surfaces with
randomly arranged active sites, then the surfaces with the stripes of
active sites of increasing width were considered. The employment of
stripes let us control the ordering of the active centers
arrangement. In each case the ratio of the number of active to
inactive sites was kept constant ($\theta_{active} = 0.5$). The same
surface models were used in our previous Kinetic Monte Carlo study
\cite{cwiklik_ssci}.

To make possible the direct comparison of results obtained using the
present model with the results of Monte Carlo simulations the
parameters of the former (the constants in Eqs.
(\ref{eq:uni_solution_x}) -- (\ref{eq:uni_solution_z}) and in the
solution of Eqs. (\ref{eq:bi_dyn_1}) and (\ref{eq:bi_dyn_2})) were
chosen in a correspondence to the parameters used in the simulations
\footnote{In the Monte Carlo simulation algorithm different number of
  processes was possible for different lattice sites and hence the
  probabilities of processes were varying between active and inactive
  centers. In order to include proper values of parameters the
  probabilities were rescaled in the following way (primed values were
  actually put into equations whereas in the text the values
  referencing to the unprimed parameters are given): $k'_1 = x_{gas}$,
  $k'_2 = {1\over3} p_{des}$, $k'_3 = x_{gas}$, $k'_4 = {1\over2}
  p_{des}$, $k'_5 = {1\over8} p_{diff}$, $k'_6 = {1\over12} p_{diff}$.
  In order to use $k'_5 = k'_6$ approximation the values of $k'_5$ and
  $k'_6$ were averaged and the arithmetic mean value $k'_5 = 0.104
  p_{diff}$ was taken into account in calculations. We also, in order
  to compare TON values, renormalized active sites coverage
  $\left[A\ast\right]' = 2 \left[A\ast\right]$ in order to get the
  coverage of active sites and not the coverage of the whole
  surface.}.

The crucial step in the implementation of the present model is the
derivation of inhomogeneity parameters $w$ and $w_r$. The former
parameter is an average probability that, if one considers a pair of
neighboring centers, an active-inactive pair is found. In order to
derive $w$ the neighborhood has to be defined. Here we consider 8
nearest lattice sites to a given one as its neighborhood (this is a
minimum neighborhood that distinguish among considered arrangements of
active sites because an alternative, 4 nearest lattice sites
neighborhood, gives $w$ that does not differ for 'rand' and '1+1'
systems). For the random distribution of active sites (and for
$\theta_{active} = 0.5$, like in the present case) $w$ is equal to 0.5
because the probability, that an active center has an active
neighbor, is 0.5 since they are spread randomly.  For '1+1' system,
$w = 6/8$ because among 8 neighbors of each active site 6 of them are
inactive.  For the systems: '2+2', '3+3' and '4+4' the values of $w$
are, respectively, equal to: $3/8$, $2.0/8$, $1.5/8$. In the two
latter cases the neighborhoods are different for centers located
inside and on the edges of active stripes therefore the average
weighted with respect to the number of differently located sites was
taken (it is depicted by the decimal representation of the numerators
2.0 and 1.5). For '1/2' system in the limit of infinite lattice $w
\rightarrow 0$ because the number of active-inactive neighbors is
negligible in comparison with the total number of pairs. According to
the values of $w$ the considered systems can be ordered with the
increasing 'inhomogeneity' as the following: '1/2', '4+4', '3+3',
'2+2', 'random', '1+1'. It is worth to note that '1+1' system is ,
according to our definition of 'inhomogeneity', less ordered than the
system with the random distribution of active centers.

The derivation of the parameter $w_r$ is less straightforward.
Considering the bimolocular reaction, we assume that 4 nearest lattice
sites form a site's neighborhood (the same assumption was taken in
the Monte Carlo simulation algorithm). The rate of bimolecular surface
reaction, from the point of view of a single active site, is
proportional to the number of active centers in the neighborhood.
Because of this it seems, that the more active neighbors are present,
the higher the rate should be. However, active centers in the
neighborhood are also competitors of the considered active site in the
usage of the reactant and due to this effect the rate is, at the same
time, inversely proportional to the number of active centers in the
neighborhood.  Therefore there exists a balance between the
cooperative and competitive aspects of active-active neighborhood. On
the other hand, at least one active neighbor is necessary for a given
active center to be able to generate product, and hence 1 active site
in the neighborhood is a threshold for reaction, and the further
increasing of the number of active neighbors above 1 does not lead to
the increase of the reaction rate.  Taking into account these
considerations, $w_r$ can be approximated as equal to the average
probability of finding at least one active center in the nearest
neighborhood of an active site (note that according to this
definition, $w_r$ does not increase when the number of active
neighbors increases above 1). The parameter $w_r$ was equal to 1 for
each surface model, except the one with randomly placed active
centers, where (in the case of 4 nearest sites neighborhood):
\begin{equation}
  w_r = 1 - (1 - \theta_{act})^4
\end{equation}
and in this system it was equal to 0.9375 for $\theta_{act} = 0.5$. In
the above equation the $(1 - \theta_{act})^4 = \theta_{inactive}^4$
term gives the probability that each of 4 neighbors is inactive one,
thus the $1 - (1 - \theta_{act})^4$ term corresponds to the
probability that at least one of neighbors is active.

\begin{figure} [!hbp]
\centering
\includegraphics[width=0.4\textwidth]{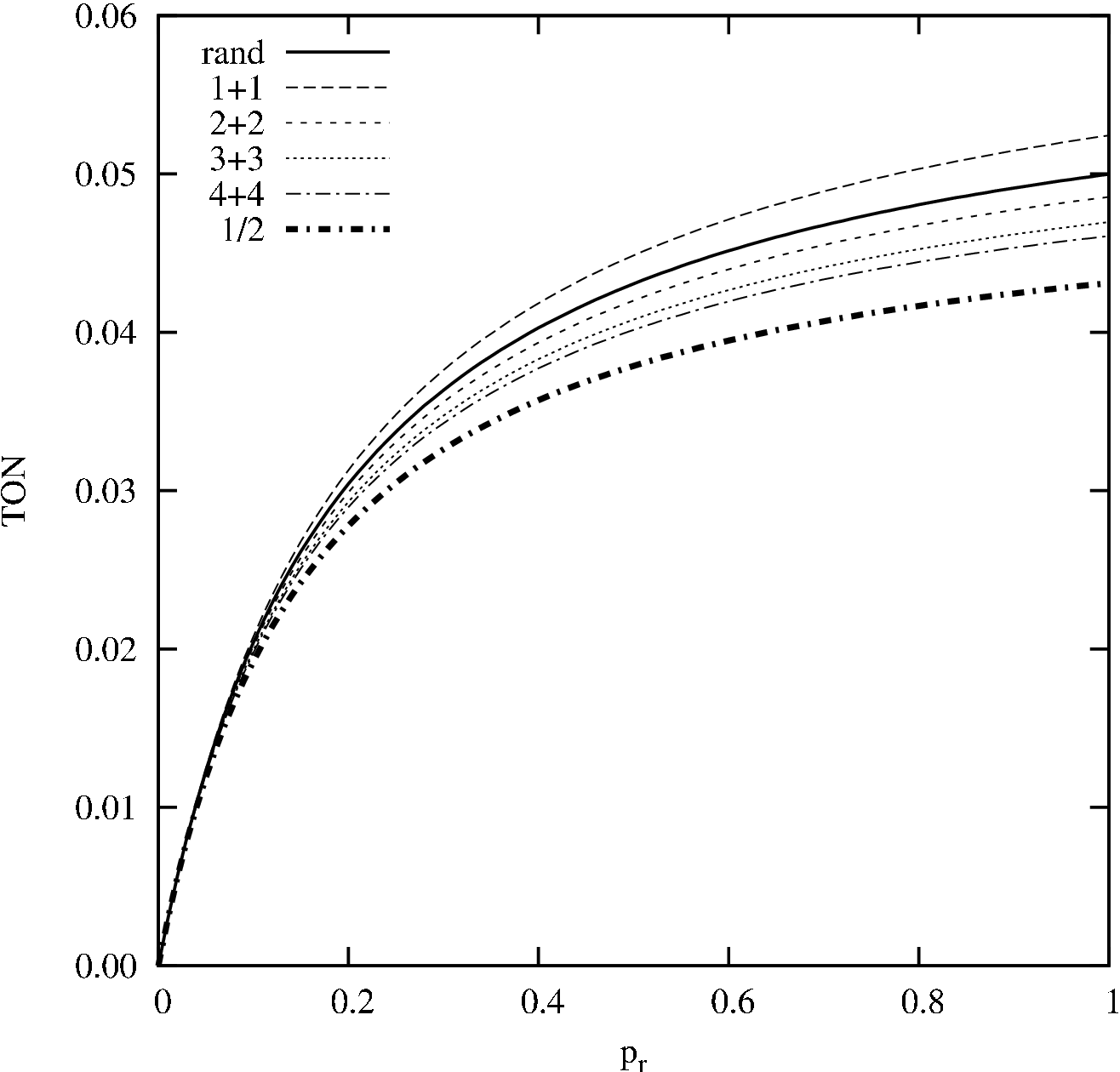}
\caption{TON vs. reaction probability for the unimolecuar surface
reaction mechanism.}
\label{fig:uni}
\end{figure}

Fig. \ref{fig:uni} shows TON vs. $p_r$ parameter for the system with
the unimolecular surface reaction mechanism. TON is calculated
employing presented analytical model with the set of model parameters
corresponding to the Monte Carlo study ($\theta{act} = 0.5$, $x_{gas}
= 0.05$, $p_{des} = 0.01$, $p_{diff} = 0.5$). Curves for various
arrangements of active centers, i.e., various values of $w$ parameter,
are presented.  The results are in good agreement with the
corresponding dependencies showed in \cite{cwiklik_ssci}. The highest
values of TON are reached for '1+1' system then the catalytic
efficiency decreases along with the increasing ordering in the
arrangement of active sites. TON for '1/2' case is, consequently, the
lowest one. There is a minor difference between Monte Carlo and
analytical model: in the latter one TON of 'random' system is nearer
TON values for '2+2' system, whereas it was more similar to the '1+1'
case in the Monte Carlo study.  Also, the values of TON obtained with
the presented model are slightly lower than those predicted in
simulations.

\begin{figure} [!hbp]
\centering
\includegraphics[width=0.4\textwidth]{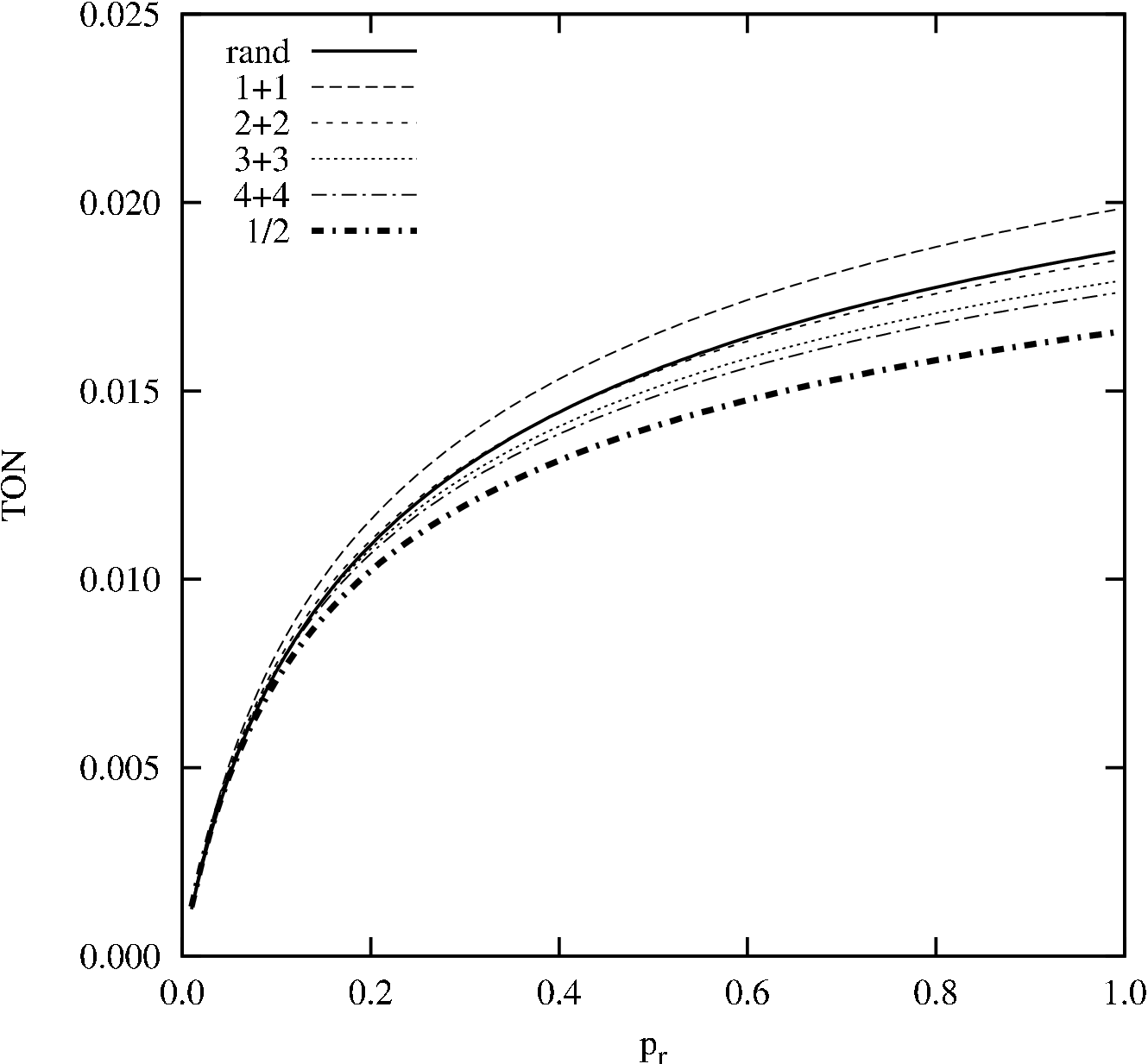}
\caption{TON vs. reaction probability for the bimolecular surface
reaction mechanism.}
\label{fig:bi}
\end{figure}

In Fig. \ref{fig:bi} the dependencies of TON on $p_r$ for systems with
the bimolecular surface reaction are presented (for the same set of
parameters as in the unimolecular case). These results are in
qualitative agreement with our previous Monte Carlo study, i.e., like
in the unimolecular case, TON decreases along with the increasing of
ordering of active centers on the surface. The system with randomly
arranged active centers is an exception, because its efficiency is
comparable with '2+2' system, whereas TON was near '4+4' system in the
results obtained in simulations. From the quantitative point of view,
the values of TON for the bimolecular reaction mechanism are lower in
comparison with those obtained in simulations.  The reasons of this
differences between the mean field model and Monte Carlo simulations
lay, most probably, in the derivation of $w_r$ parameter wherein the
size of considered neighborhood can be chosen in different ways. Since
in this work we are mainly interested in the qualitative description
of considered catalytic systems, we may conclude that in both
unimolecular and bimolecular case the mean field model is able to
reproduce the most important features of deliberated systems.

\newpage

\begin{thebibliography}{00}



\bibitem {sam2006} G.A. Samorjai, K.M. Bratlie, M.O. Montano,
 J.Y. Park, J. Phys. Chem. B 110 (2006) 20014.

\bibitem{jansen1999} A.P.J. Jansen, C.G.M. Hermse, Phys. Rev. Lett. 83
(1999) 3673.

\bibitem {zhdanov2002} V.P. Zhdanov, Surf. Sci. 500 (2002) 966.

\bibitem {temel2007} B. Temel, H. Meskine, K. Reuter, M. Scheffler,
H. Meitu, J. Chem. Phys. 126 (2007) 204711.

\bibitem {chonkbook} I. Chorkendorf, J.W. Niemantsverdriet, Concepts
of Modern Catalysis and Kinetics, Wiley-VCH, Weinheim, 2003.

\bibitem{ziff1986} R.M. Ziff, E. Gulari, Y. Barshad,
Phys. Rev. Lett. 56 (1986) 2553.

\bibitem{hermse2000} C.G.M. Hermse, A.P.J. Jansen, Surf. Sci. (2000)
168.

\bibitem{mcleod2000} A.S. McLeod, L.F. Gladden,
J. Chem. Inf. Comput. Sci. 40 (2000) 981.

\bibitem{barnes2001} R. Barnes, I.M. Abdelrehim, T.E. Madey,
Top. Catal. 14 (2001) 53.

\bibitem{tammaro1995} M. Tammaro, M. Sabella, J.W. Evans,
J. Chem. Phys. 103 (1995) 10277.

\bibitem{james1999} E.W. James, C. Song, J.W. Evans,
J. Chem. Phys. 111 (1999) 6579.

\bibitem{evans2002} J.W. Evans, D.-J. Liu, M. Tammaro, Chaos 12 (2002)
131.

\bibitem{liu2002} D.-J. Liu, J.W. Evans, J. Chem. Phys. 117 (2002)
7319.

\bibitem{tammaro1998} M. Tammaro, J.W. Evans, J. Chem. Phys. 108
(1998) 762.

\bibitem{liu2006} D.-J. Liu, J.W. Evans, J. Chem. Phys. 125 (2006)
054709.

\bibitem{mastny2006} E.A. Mastny, E.L. Haseltine, J.B. Rawlongs,
J. Chem. Phys. 125 (2006) 194715.

\bibitem{caballero2005} F.V. Caballero, L. Vincente, Chem. Eng. J. 106
(2005) 229.


\bibitem{cwiklik_spill} L. Cwiklik, B. Jagoda-Cwiklik, M. Frankowicz,
Appl. Surf. Sci. 252 (2005) 778.

\bibitem{hu2005} R. Hu, S. Huang, Z. Liu, W. Wang,
Appl. Surf. Sci. 242 (2005) 353.

\bibitem{zvejnieks2002} G. Zvejnieks, V.N. Kuzukov, Phys. Rev. E 66
(2002) 021109.

\bibitem{dickman1986} R. Dickman, Phys. Rev. A 34 (1986) 4246.

\bibitem{cwiklik_ssci} L. Cwiklik, B. Jagoda-Cwiklik, M. Frankowicz,
Surf. Sci. 572 (2004) 318.

\bibitem{libuda2005} J. Libuda, H.-J. Freund, Surf. Sci. Rep. 57
(2005) 157.


\end{thebibliography}
\end{document}